\documentstyle [aps,preprint]{revtex}
\tightenlines
\begin{document}
\def\fnote#1#2{
\begingroup\def\thefootnote{#1}\footnote{#2}\addtocounter{footnote}{-1}
\endgroup}
\def\dslash{\not{\hbox{\kern-2pt $\partial$}}}
\def\eslash{\not{\hbox{\kern-2pt $\epsilon$}}}
\def\Dslash{\not{\hbox{\kern-4pt $D$}}}
\def\Aslash{\not{\hbox{\kern-4pt $A$}}}
\def\Qslash{\not{\hbox{\kern-4pt $Q$}}}
\def\Wslash{\not{\hbox{\kern-4pt $W$}}}
\def\pslash{\not{\hbox{\kern-2.3pt $p$}}}
\def\kslash{\not{\hbox{\kern-2.3pt $k$}}}
\def\qslash{\not{\hbox{\kern-2.3pt $q$}}}
\def\np#1{{\sl Nucl.~Phys.~\bf B#1}}
\def\pl#1{{\sl Phys.~Lett.~\bf B#1}}
\def\pr#1{{\sl Phys.~Rev.~\bf D#1}}
\def\prl#1{{\sl Phys.~Rev.~Lett.~\bf #1}}
\def\cpc#1{{\sl Comp.~Phys.~Comm.~\bf #1}}
\def\cmp#1{{\sl Commun.~Math.~Phys.~\bf #1}}
\def\anp#1{{\sl Ann.~Phys.~(NY) \bf #1}}
\def\etal{{\em et al.}}
\def\half{{\textstyle{1\over2}}}
\def\halfhalf{{\textstyle{1\over4}}}
\def\be{\begin{equation}}
\def\ee{\end{equation}}
\def\ba{\begin{array}}
\def\ea{\end{array}}
\def\tr{{\rm tr}}
\def\Tr{{\rm Tr}}
\title{Quantization of maximally-charged slowly-moving black holes
\thanks{Research
supported by the DoE under grant DE--FG05--91ER40627.}}
\author{
George Siopsis
\fnote{\ddagger}{E-mail: \tt gsiopsis@utk.edu}}
\address{Department of Physics and Astronomy, \\
The University of Tennessee, Knoxville, TN 37996--1200.\\
}
\date{August 2000}
\preprint{UTHET--00--0801}
\maketitle
\begin{abstract}
We discuss the quantization of a system of slowly-moving
extreme Reissner-Nordstr\"om black holes. In the near-horizon limit, this
system has been shown to possess an $SL(2, {\bf R})$ conformal symmetry.
However, the Hamiltonian appears to have no well-defined ground state. This
problem can be circumvented by a redefinition of the Hamiltonian due to de
Alfaro, Fubini and Furlan (DFF). We apply the Faddeev-Popov
quantization procedure to show that the Hamiltonian with no ground state
corresponds to a gauge in which there is an obstruction at the
singularities of moduli space requiring a modification of the
quantization rules. The redefinition of the Hamiltonian {\em \`a la} DFF
corresponds to a different choice of gauge. The latter is a good gauge
leading to standard quantization rules. Thus, the DFF trick is a consequence
of a standard gauge-fixing procedure in the case of black hole scattering.
\end{abstract}
\renewcommand\thepage{}\newpage\pagenumbering{arabic} 
\section{Introduction} 
The study of moduli space of a system of maximally charged black holes
has recently attracted a lot of attention~\cite{bib4}. It was first discussed
by Ferrell and Eardley~\cite{bib5} in four spacetime dimensions. It was
subsequently extended to five dimensions~\cite{bib4}. In the near-horizon
limit, an $SL(2, {\bf R})$ conformal symmetry was discovered that generalized
the case of two black holes. Consequently, the general system inherited the
pathologies of the system of two black holes: the Hamiltonian possessed no
well-defined vacuum state.

This problem was studied a long time ago by de Alfaro, Fubini and Furlan
(DFF)~\cite{bib1}. The simplest quantum mechanical system with conformal
symmetry is described by the Hamiltonian
\be H = {p^2\over 2} + {g\over 2x^2} \ee
This Hamiltonian represents a single-particle rational Calogero-Moser system,
which is equivalent to a free particle system~\cite{bib6}. It possesses a
continuous spectrum down to zero energy and there is no well-defined ground
state. DFF suggested a solution to this problem. They proposed the
redefinition of the Hamiltonian by the addition of a harmonic oscillator
potential which is also the generator of special conformal transformations,
\be K = {x^2\over 2} \ee
The new Hamiltonian is defined by
\be H' =
{1\over \omega} (H+\omega^2 K)\ee
where we introduced a scale parameter
$\omega$ (infrared cutoff). $H'$ has a well-defined vacuum and a discrete
spectrum, which can actually be computed exactly, 
\be E_n =
2n+1+\sqrt{g+1/4} \ee
Notice that the spectrum is independent of the arbitrary scale parameter 
$\omega$.
The supersymmetric case can be dealt with in the same way.

In the case of two slowly-moving maximally-charged black
holes,~\cite{bib2,bib2a,bib2b,bib2c,bib3}. the near-horizon geometry is
$AdS_2\times S^n$. The isometries of the $AdS_2$ space are conformal
symmetries. As a result, the moduli (spatial distance between the two black
holes) has dynamics governed by (super)conformal non-relativistic quantum
mechanics of the form discussed above. The DFF
redefinition of the Hamiltonian has a nice interpretation in this case as a
redefinition of the time coordinate. The DFF Hamiltonian corresponds to a
globally defined time coordinate whereas the conformally invariant definition
does not. Thus, the DFF trick appears plausible on physical grounds.

The multiple black hole moduli space possesses similar properties~\cite{bib4}.
There is a conformal $SL(2,{\bf R})$ symmetry and the quantization of the
system leads to a Hamiltonian with no well-defined ground state. The DFF trick
of replacing the Hamiltonian $H$ by $H+K$, where $K$ is the generator of
special conformal transformations, is applicable in this general case.
However, it lacks an obvious physical interpretation in terms of a
redefinition of the time coordinate.

In ref.~\cite{bib7}, we presented an alternative derivation of the DFF
procedure in the case of a particle moving in the background of an extreme
Reissner-Nordstr\"om black hole. This is equivalent to a system of two black
holes. We showed that the redefinition of the Hamiltonian amounted to a
different choice of gauge. In the conformally invariant case, we identified an
obstruction to the standard gauge-fixing procedure that led to a
modification of the usual quantization rules. This obstruction came from
the boundary of spacetime and was rooted in the fact that the time coordinate
was not defined at the boundary. On the other hand, there was no obstruction
in the choice of gauge leading to the DFF Hamiltonian. We concluded that the
DFF Hamiltonian corresponded to a good gauge choice, whereas the conformally
invariant Hamiltonian did not. Our discussion was based on the standard
Faddeev-Popov quantization procedure and was therefore applicable to more
general systems, as long as the system had an underlying gauge invariance.

Here, we extend the procedure discussed in~\cite{bib7} to the case of
multiple black hole scattering~\cite{bib4}. We show how the gauge can be fixed
systematically without encountering obstructions from the singularities of
moduli space. The resultant Hamiltonian is modified by the addition of the
potential prescribed by the DFF trick. Thus, we show that the DFF trick is a
consequence of a standard gauge-fixing procedure in the case of multiple black
hole scattering. 

Our discussion is organized as
follows. In Section~\ref{sec2}, we apply the Faddeev-Popov procedure to a
particle moving in a fixed background of curved spacetime as well as an
external electromagnetic field. We also show how the procedure is
equivalent to the commutation rules one obtains from Dirac brackets.
We discuss a subtlety that arises when spacetime possesses boundaries. In
Section~\ref{sec4}, we specialize to the case of an extreme
Reissner-Nordstr\"om black hole in five spacetime dimensions. We show that
the DFF trick is equivalent to a choice of gauge. In Section~\ref{sec4a}, we
discuss multiple black hole scattering and show how the system of black holes
can be quantized leading to a modification of the Hamiltonian {\em \`a la} DFF.
In Section~\ref{sec4D}, we discuss the case of four spacetime dimensions.
Finally, in Section~\ref{sec5}, we
summarize our conclusions.

\section{Charged particle in curved
spacetime} \label{sec2}  In this Section, we discuss the quantization of a
charged particle moving in a fixed spacetime background and electromagnetic
field. We introduce the path integral in curved spacetime and apply
the Faddeev-Popov procedure to fix the gauge. We also show that this is
equivalent to the canonical quantization through commutation relations coming
from Dirac brackets. We discuss a subtlety that arises in the quantization
procedure when spacetime has boundaries. This is a review of
ref.~\cite{bib7}.  Consider a particle of mass $m$ and charge $q$ moving
along a trajectory described by coordinates $x^\mu (\tau)$ ($\mu = 0,1,\dots
,D-1$) where $\tau$ is the proper time of the particle. The action is
\be\label{chaction0}
S = \int d\tau \; L\quad, \quad L = {1\over 2\eta} \dot
x^\mu \dot x_\mu -\half\eta m^2 + q \dot x^\mu A_\mu \ee
where we raise and
lower indices with the background metric $g_{\mu\nu}$. $A_\mu$ is the
background electromagnetic vector potential. Varying $\eta$, we obtain the
constraint 
\be
\eta^2 = - \dot x^\mu \dot x_\mu / m^2 \ee The conjugate
momenta are 
\be
P_\mu = {\partial L\over \partial \dot x^\mu} =
{1\over\eta}\; \dot x_\mu + q A_\mu \quad,\quad P_\eta = 0 \ee The
Hamiltonian is 
\be
H = \dot x^\mu P_\mu - L = m\eta \chi\quad,\quad \chi =
{1\over 2m} \, \pi_\mu \pi^\mu + \half m\quad, \quad \pi_\mu = P_\mu-qA_\mu
\ee In the canonical formalism, the action reads 
\be\label{chaction}
S =
\int d\tau \left( \dot x^\mu P_\mu - m\eta \chi \right) \ee
Therefore, $\eta$ is a Lagrange multiplier enforcing the constraint
\be\label{constraint2}
\chi\equiv {1\over 2m} \, \pi_\mu \pi^\mu + \half m = 0
\ee
which is the mass-shell condition in the presence of an external vector 
potential. 

The orbits of the gauge transformations ($\tau$ reparametrizations) are the
trajectories of the equations of motion (Lorentz force law in curved spacetime)
\be
\dot x^\mu = {1\over m}\, \pi^\mu \quad,\quad \dot \pi_\mu + {1\over m} 
\Gamma_{\nu\lambda\mu} \pi^\nu\pi^\lambda = {q\over m} \, \pi^\nu F_{\mu\nu}
\quad,\quad F_{\mu\nu} = \partial_\mu A_\nu
- \partial_\nu A_\mu
\ee
where $\Gamma_{\nu\lambda\mu}$ are the Christoffel symbols,
\be
\Gamma_{\nu\lambda\mu} = \half(\partial_\lambda g_{\mu\nu} - \partial_\nu
g_{\lambda\mu} + \partial_\mu g_{\nu\lambda})
\ee
or purely in terms of the coordinates $x^\mu$,
\be\label{eqcoo}
\ddot x^\mu + \Gamma_{\nu\lambda}^\mu \dot x^\nu \dot x^\lambda = {q\over m} \, 
\dot x^\nu F_{\;\;\nu}^\mu
\ee
To quantize the system, consider the path integral,
\be
Z = {\cal N} \int {\cal D} x {\cal D} P {\cal D} \eta \; e^{iS}
= {\cal N} \int {\cal D} x {\cal D} P\; \delta(\chi)\; e^{i\int d\tau\,
\dot x^\mu P_\mu}
\ee
To define it, we need to fix the gauge
by imposing the gauge-fixing condition
\be\label{kons}
h(x^\mu) = \tau
\ee
which defines a hyper-surface that cuts each orbit precisely once.
Physically, this amounts to choosing $h(x^\mu)$ as the time coordinate.
Then its conjugate momentum, ${\cal H}$, is the Hamiltonian of the reduced
system.
Following the standard Faddeev-Popov procedure, we insert
\be\label{fpdet}
1 = \det \{h,\chi\}\; \int {\cal D}\epsilon  \; \delta (h - \{h,\chi \}\epsilon - 
\tau)
\ee
where $\{\;,\;\}$ denotes Poisson brackets,
into the path integral and perform a reparametrization to obtain
\be
Z = {\cal N} \int {\cal D} x {\cal D} P\; \det \{h,\chi\}\;
\delta (h - \tau)\; \delta(\chi)\; e^{i\int d\tau\,
\dot x^\mu P_\mu}
\ee
We may integrate over the $\delta$-functions to reduce the dimension of phase 
space.
The reduced system will be described by coordinates $\overline x^i$ and
conjugate momenta $\overline P_i$. The Faddeev-Popov determinant is canceled by
the integration over $\delta(\chi)$. The momentum conjugate to $h$ (which is
identified with time) plays the
r\^ole of the Hamiltonian ${\cal H}$ of the reduced system. The path integral
becomes
\be
Z = {\cal N} \int {\cal D} \overline x {\cal D} \overline P\; e^{i\int d\tau\,
(\dot{\overline x}^i \overline P_i -{\cal H})}
\ee
Equivalently, we may quantize the system in the operator formalism. To this end, 
we need to calculate Dirac brackets,
\be
\{\, A\;,\; B \,\}_D = \{\, A\;,\; B \,\} - \{\, A\;,\; \chi_i \,\}\;
\{\, \chi_i\;,\; \chi_j \,\}^{-1}\; \{\, \chi_j\;,\; B \,\}
\ee
where $i,j=1,2$, $\chi_1 = \chi$, $\chi_2 = h$, and promote them to commutators.

As an example, consider the special case $h(x^\mu) = x^0$ ({\em i.e.,}
identify $x^0$ with time), and set $q=0$. The reduced system is described by the 
coordinates
$\overline x^i = x^i$ and the Hamiltonian is
\be\label{hp0}
{\cal H} = -P_0 = \sqrt{P_iP^i+m^2}
\ee
%Then $\partial_\pm h = 1$ and
%\be H=P_++P_- = P_0\ee
The commutation relations we obtain from the Dirac brackets are
%in terms of the independent coordinates $x^1$, $x^i$ are
\be
[ P_i \;,\; x^j ] = -i \delta_i^{\; j}
\quad,\quad [{\cal H}\;,\; x^i ] = -i\, {P^i\over {\cal H}}
\ee
which are appropriate for ${\cal H}$ given by (\ref{hp0}).

Having set up the quantization procedure, we now wish to discuss a subtlety 
which
arises when the spacetime possesses boundaries~\cite{bib8}.
Let us follow the Faddeev-Popov gauge-fixing procedure a
little more carefully. We need to insert~(\ref{fpdet}) into the path integral
and then perform an inverse gauge transformation to eliminate the gauge 
parameter.
In doing so, we encounter an obstruction at the boundary of spacetime.
Under a gauge transformation, the change in the action is
\be
\delta S = \int d\tau \; {d\over d\tau} (\delta x^\mu P_\mu) - \int d\tau \,
\epsilon \dot\chi
\ee
Since $\dot\chi = \{\chi \,,\, \chi\} =0$, we conclude that
the action changes by a total derivative,
\be
\delta S = \int d\tau \; {d\over d\tau} (\delta x^\mu P_\mu)
\ee
We have
\be
\delta x^\mu = \{ x^\mu\,,\, \chi\} \, \epsilon = {\partial\chi\over \partial 
P_\mu} \, \epsilon
\ee
therefore,
\be
\delta S = P_\mu {\partial\chi\over \partial P_\mu} \, \epsilon \, 
\Bigg|_\partial
\ee
Notice that, if the generator of gauge transformations, $\chi$,
is quadratic in the momenta (as in the neutral particle
case), then the boundary contribution vanishes after imposing the constraint
$\chi =0$. In general, $\chi$ (Eq.~(\ref{constraint2})) is not quadratic in the 
momenta, due to the
presence of the vector potential. Therefore $\delta S\ne 0$ and
we cannot in general get rid of the gauge parameter on the boundary
of spacetime. Thus, we obtain
a boundary contribution to the path integral,
\be
\int_\partial d\epsilon d^Dx d^DP \, \{\, h\,,\, \chi \,\}
\, \delta (h-\tau) \delta(\chi)\, \exp\left\{ iP_\mu {\partial\chi\over \partial 
P_\mu} \epsilon \right\}
\ee 
This obstruction is absent when at the boundary,
\be \{\, h\,,\, \chi \,\}\Big|_\partial =0\ee
Physically, this condition implies that the boundary of spacetime is invariant 
under transformations
generated by $h$, which is the time coordinate after gauge-fixing ($h=\tau$).
In other words, the boundary is fixed under time translations. Thus the
time coordinate $h$ is not a good global coordinate and leads to an
obstruction in the gauge invariance of the theory. Integrating over the gauge
parameter, we obtain an additional constraint at the boundary,
\be
P_\mu {\partial\chi\over \partial P_\mu}\; \Bigg|_\partial = 0
\ee
This alters the standard commutation relations and the eigenvalue problem for
the Hamiltonian. We have not carried out an explicit
computation. This would involve the introduction of a regulator which would 
break
gauge invariance. Nevertheless, the resulting system should be equivalent to the
one obtained through other choices of gauge due to the gauge invariance of the
theory.

To summarize, the identification of the time coordinate~(\ref{kons}) in 
general
leads to an obstruction in the gauge-fixing procedure for the path integral.
If this obstruction is accounted for by an appropriate modification of the
commutation relations, this choice of the time coordinate leads to a well-
defined
Hamiltonian problem.

%\vspace{0.5cm}\hrule\vspace{2pt}\hrule\vspace{0.5cm}

\section{Extreme Reissner-Nordstr\"om black hole}
\label{sec4}

We are now ready to discuss the quantization of a particle moving near an 
extreme
Reissner-Nordstr\"om black hole~\cite{bib2,bib2a,bib2b,bib2c,bib3}. In this
Section, we will discuss five spacetime
dimensions. The results in the four-dimensional case are similar and will be
taken up in Section~\ref{sec4D}.

Consider a maximally charged black hole sitting at the origin of spacetime.
We shall work with units in which Newton's constant $G=1$. Then
\be M = Q \ee
for this black hole. The particle moving near the black hole will also be
taken
as
 maximally charged, so
\be\label{eqmq} m =  q \ee
The black hole creates a metric
\be
ds^2 = - {1\over \psi^2} dt^2 + \psi d\vec x^2 \quad,\quad \psi = 1 +
{M\over \vec x^2}
\ee
and a vector potential
\be
A_0 = {1\over\psi} \quad,\quad \vec A = 0
\ee
where the vectors live in a four-dimensional Euclidean space.
Near the horizon, $\psi = M/\vec x^2$. Using polar coordinates and switching
variables to $\psi$, we obtain
\be
ds^2 = -{1\over\psi^2} \left( dt^2 - {M\over 4}\, d\psi^2\right)
+ M d\Omega_3^2
\ee
Defining
\be
\label{eq34}
x^\pm = t \pm {\sqrt M\over 2} \; \psi
\ee
the metric becomes
\be\label{metr4}
ds^2 = - {1\over \psi^2} \; dx^+ dx^- + M d\Omega_3^2
\ee
and
\be
\psi = {x^+ - x^-\over \sqrt M}
\ee
The vector potential has non-vanishing components
\be
A_+ = A_- = {1\over 2\psi}
\ee
Thus,
spacetime factorizes into a product $AdS_2\times S^3$.
Henceforth, we shall work
with $AdS_2$.
The only non-vanishing connection coefficients are $\Gamma_{\pm\pm}^\pm =
\partial_\pm \ln |g_{+-}|$. Therefore,
the geodesic equations for $x^\pm$ are (setting $m=q$ (Eq.~(\ref{eqcoo})))
\be
\ddot x^\pm \pm  (\ln |g_{+-}|)' (\dot x^\pm)^2 = \pm \; \dot x^\pm F_{+-}
\ee
where $A=A_+=A_-$, and $(\ln |g_{+-}|)' = \partial_+ \ln |g_{+-}| = - \partial_- 
\ln |g_{+-}|$.
%It is easy to see that the orbits with constant $\psi$ are geodesics,
%\be
%x^+ = x^-+ \sqrt M\alpha
%\ee
%where $\psi = \alpha$, parametrized by
%\be
%\tau = 2\alpha \, (x^++x^-)
%\ee
%Motion along these geodesics is generated by the conjugate momentum,
%\be H = -P_0 = -P_+-P_-\ee
%
%Another interesting set of geodesics can be obtained by changing variables to
%\be
%\theta^\pm = \arctan (x^\pm /Q)
%\ee
Using $\psi\, {dA\over d\psi} = -A$, $\psi\, {dg_{+-}\over d\psi} = - 2g_{+-}$,
$F_{+-} = 2\partial_+A$,
it is straightforward to show that the following quantities are gauge-invariant
(constant along geodesics)
\be\label{eq42}
H = -P_+-P_-\quad,\quad D = 2x^+P_++2x^-P_-\quad,\quad
K = -(x^+)^2P_+ - (x^-)^2P_- + \half mM\psi
\ee
They obey an $SL(2,{\bf R})$ algebra
\be
\{\,H\;,\; D\,\} = -2H\quad,\quad \{\,H\;,\;K\,\} =-D\quad,\quad
\{\,K\;,\;D\,\} = 2K
\ee
reflecting the symmetry of the $AdS_2$ spacetime.
$H, D$, and $K$ generate time translations, dilatations, and special conformal
transformations, respectively.
The brackets may be Poisson or Dirac, so this is also an algebra of the
gauge-fixed system, as expected.
The constraint (generator of gauge transformations)
$\chi\equiv {1\over 2m} \pi_\mu \pi^\mu + {1\over 2} m = 0$ reads
\be\label{eq44}
2m \chi = - \psi^2 P_+P_- +\half m\psi (P_++P_-) +{L^2\over M}  =0
\ee
where $L^2 = \hat g^{ij} P_iP_j$ is the square of the angular momentum operator.
The simplest gauge-fixing condition to impose is
\be\label{gauge1}
h(x^+,x^-) = \half (x^++x^-) =\tau
\ee
In this case, the Hamiltonian is
\be\label{eq46}
H = -P_+-P_-
\ee
Using the constraint~(\ref{eq44}), we obtain
\be\label{eq54}
H = {1\over \psi} \, \left(-m+\sqrt{m^2+ 4(\psi^2P_\psi^2+L^2)/M}\right)
\ee
The other two operators in the $SL(2,{\bf R})$ algebra can be written as
\be
D = -2\tau H+2\psi P_\psi \quad,\quad K = \half \tau^2 H-\halfhalf\tau D+
\halfhalf M\psi^2 H +\half mM\psi
\ee
In the non-relativistic limit and for large $\psi$ (near the horizon),
\be
H = {2\psi P_\psi^2\over mM}\quad,\quad
D = 2\psi P_\psi \quad,\quad K=  \half mM\psi
\ee
where we also used the fact that these are conserved quantities to set
$\tau =0$.
Switching back to $\vec x$, we may write these operators in terms of the
coordinate $\vec x$ and its conjugate momentum $\vec P$ as
\be\label{eq50}
H = {\vec x^4 \, \vec P^2\over 2mM^2} \quad,\quad
D = - \vec x\cdot \vec P \quad,\quad K = {mM^2\over 2\, \vec x^2}
\ee
In terms of a variable $u$,
defined by
\be
\psi = {u^2\over M}
\ee
we obtain a simple representation,
\be
H = {P^2\over 2m} \quad,\quad
D = u P \quad,\quad K= \half m u^2
\ee
where $P$ is the momentum conjugate to $u$.
This system does not have a well-defined vacuum.
The question then arises whether the underlying theory is inherently sick.
One may apply the DFF trick to produce a Hamiltonian system with a well-defined
ground state. The DFF trick can be understood in this case as a different choice 
of
time coordinate leading to a different Hamiltonian.
From our point of view, any two choices of time coordinates should be equivalent 
to
each other, for they merely correspond to different gauge choices.
Since the underlying theory is gauge-invariant, all gauge choices should be
equivalent to each other.

Before we discuss the vacuum problem in conjunction with the gauge-fixing 
procedure,
we shall introduce a class of gauges that lead to a Hamiltonian system with a
well-defined ground state.
Let the gauge-fixing condition be
\be\label{gftan}
h(x^+,x^-) = \arctan \left( {\omega x^++\omega x^-
\over 1-\omega^2 x^+x^-} \right) = \tau
\ee
where $\omega$ is an arbitrary scale. Differentiating with respect to $\tau$,
we obtain
\be
\partial_+ h \, \dot x^+ + \partial_- h \, \dot x^- =1\quad,\quad
\partial_\pm h = {\omega\over 1+\omega^2 (x^\pm)^2}
\ee
To find the Hamiltonian, we start from the Lagrangian,
\be\label{eq55}
L = \dot x^+ P_+ + \dot x^- P_- + \dot \Lambda
\ee
where we added the time derivative of a function to be specified shortly. This
does not alter the dynamics, provided there is no boundary contribution.
Alternatively, it can be viewed as a gauge transformation ($A \to A + d\Lambda$).
By introducing the coordinate
\be
\zeta = \arctan \left( {\omega x^+-\omega x^-
\over 1+\omega^2 x^+x^-} \right)
\ee
we can write the Lagrangian as
\be
L = \dot \zeta P_\zeta - \dot h {\cal H}
\ee
where
\be\label{eq58}
{\cal H} = - {1\over 2} \left( {P_+\over\partial_+ h} + {P_-\over\partial_- h}
\right) -\partial_h
\Lambda = {1\over 2\omega}
\, (H + \omega^2 K')\quad, \quad K' = -(x^+)^2P_+ - (x^-)^2P_- - {2\over
\omega} \, \partial_h \Lambda
\ee
is the momentum conjugate to $h$ and $P_\zeta$ is conjugate to $\zeta$. Since
$\dot h =1$, it follows that ${\cal H}$ plays the role of the Hamiltonian.
We will choose $\Lambda$ so that $K' = K$~(Eq.~(\ref{eq42})). This ensures that the constraint
$\chi$ (Eq.~(\ref{eq44})) and therefore the Hamiltonian will have no explicit
time dependence, because
\be
\partial_h\chi = \{\chi\;,\; K\} = 0
\ee
due to the conservation of the charge $K$. It is easy to see that
\be\label{eq59}
\Lambda = {m\sqrt M\over 4} \; \ln {\partial_+ h\over \partial_-h}
= - {m\sqrt M\over 4} \; \ln {1+\omega^2 (x^+)^2\over 1+\omega^2 (x^-)^2}
\ee
To obtain the Hamiltonian, we solve the constraint~(\ref{eq44}). The result is~({\em cf.}~Eq.~(\ref{eq54}))
\be
{\cal H} = {\sqrt M\over\sin\zeta}\; \left( -m\cos\zeta + \sqrt{m^2\cos^2\zeta
+(4\sin^2\zeta\, P_\zeta^2/M + \half m^2M\sin^2\zeta + 4L^2)/M} \right)
\ee
The non-relativistic limit can be obtained from the above expression as the limit
$\zeta\to 0$,
\be
{\cal H} = {1\over 2\omega} \; \left( {P^2\over 2m} + \half m
\omega^2 u^2\right)
\ee
where $u^2 = M\zeta \approx M\omega (x^+ - x^-)$, and $P$ is the momentum
conjugate to $u$,
which has a well-defined vacuum (harmonic oscillator). All these gauges are of 
course equivalent.
Therefore, no physical quantities should depend on the scale parameter $\omega$.
In particular, notice that the spectrum in the non-relativistic limit is
independent of $\omega$.

It should be pointed out that the non-relativistic limit can also be easily
deduced from~(\ref{eq58}),
\be\label{eq60}
{\cal H} = {1\over 2\omega} (H +\omega^2 K)\quad,\quad K \approx {2\over\omega} \partial_h\Lambda = \half mM\psi
\ee
which is the Hamiltonian in the na\"\i ve gauge~(\ref{eq46}) corrected by the
potential $K$~({\em cf.}~Eq.~(\ref{eq50}))
Hence the origin of the non-relativistic potential $K$ is the total time derivative $\dot\Lambda$ that we added to the
Lagrangian~(Eq.~(\ref{eq55})). Thus, even though the total time derivative does
not alter the dynamics, it is very useful in the calculation of the Hamiltonian
in the non-relativistic limit.

In the set of gauges~(\ref{gftan}), there is no boundary
contribution, because the Faddeev-Popov determinant vanishes there. Indeed,
\be
\{\, h\,,\, \chi \,\} = \dot h = {\omega \dot x^+\over 1 + \omega^2 (x^+)^2} +
{\omega \dot x^-\over 1 + \omega^2 (x^-)^2}
\ee
which vanishes as $x^\pm\to\infty$ for finite velocities $\dot x^\pm$.
Therefore, one obtains standard commutation relations in this gauge.
It should also be pointed out that, since we insist $\dot h = 1$,
the region near the boundary does not contribute in the non-relativistic
limit.

On the other hand, for the gauge~(\ref{gauge1}), we obtain
a boundary contribution to the path integral,
\be
\int_\partial d\epsilon d^Dx d^DP \, \{\, h\,,\, \chi \,\}
\, \delta (h-\tau) \delta(\chi)\, \exp\left\{ iP_\mu {\partial\chi\over \partial 
P_\mu} \epsilon \right\}
\ee 
Thus, the na\"\i ve identification of the time
coordinate~(\ref{gauge1}) leads to an obstruction in the gauge-fixing
procedure for the path integral. If this obstruction is accounted for by an
appropriate modification of the commutation relations, this choice of the time
coordinate leads to a well-defined Hamiltonian problem. The Hamiltonian system
thus obtained is equivalent to applying the DFF trick~\cite{bib1}, or
identifying the time coordinate as in Eq.~(\ref{gftan})~\cite{bib2c}. The
latter is merely a different gauge choice in a gauge-invariant theory.

\section{Slowly-moving Reissner-Nordstr\"om black holes}
\label{sec4a}

Having understood the quantization of a particle in a background created by a
Reissner-Nordstr\"om black hole, we now turn to a discussion of the quantization
of a system of dynamical Reissner-Nordstr\"om black holes. Again, we consider
five spacetime dimensions. The results are similar in the four-dimensional case
(see Section~\ref{sec4D}).

The action may be written as
\be
S = S_{fields} + S_{source}
\ee
where the action for the fields is
\be\label{eqA}
S_{fields} = {1\over 12\pi^2} \int d^5 x\, \sqrt g \Big(R - {\textstyle{3\over
4}} F^2\Big) + {1\over 12\pi^2} \int A \wedge F\wedge F
\ee
in terms of a dynamical metric field $g_{\mu\nu}$ and electromagnetic vector
potential $A_\mu$, both functions of the coordinates $x^\mu$ ($\mu = 
0,1,\dots,4$).
The sources are described by coordinates $X_I^\mu$, where the index $I$ labels
the source and conjugate momenta $P_{I\mu}$. The action is
\be\label{eq65}
S_{source} = \sum_I \int dX_I^\mu P_{I\mu}
\ee
together with the constraints
\be\label{eq62}
\chi_I \equiv {1\over 2M_I} \, \pi_{I\mu} \pi_I^\mu + \half M_I =0\quad,\quad
\pi_{I\mu} = P_{I\mu} - Q_I A_\mu\quad,\quad Q_I = M_I
\ee
There is also a fermionic contribution which we omit because it is of no 
relevance
to our discussion.

Using the {\em ansatz}
\be
ds^2 \equiv g_{\mu\nu}\, dx^\mu dx^\nu = - {1\over \psi^2}\, dt^2 + \psi\, d\vec 
x^2
+ 2\vec N \cdot d\vec x\, dt
\ee
\be\label{extder}
A = {1\over\psi}\, dt + \vec A \cdot d\vec x
\ee
and keeping only terms quadratic in the potentials $\vec N,\vec A$ and 
discarding total derivatives,
the action for the fields becomes~\cite{bib4}
$$S_{fields} = {1\over 12\pi^2} \int d^5 x \; \left( 3\partial_t
\vec P\cdot\vec\partial \psi - {3\over 4\psi}\, F^2 + {3\over 2\psi^2}\;
FG - {1\over 2\psi^3} \; G^2 \right. $$
\be\label{eq69}
\left. - 3\psi (\partial_t\psi)^2 - {3\over 4\psi} \; F\widetilde F +
{3\over 4\psi^2} \; F\widetilde G - {1\over 4\psi^3} \; G\widetilde G\right)
\ee
where we introduced the convenient (gauge-invariant) combinations
\be\label{eq70}
\vec P = \vec A + \psi \vec N\quad,\quad \vec R = \psi^2 \vec N
\ee
whose field strengths respectively are
\be
F_{ij} = \partial_i P_j - \partial_j P_i\quad,\quad G_{ij} = \partial_i R_j
-\partial_j R_i
\ee
and their duals: $\widetilde F^{ij} = \epsilon^{ijkl} F_{kl}$,
$\widetilde G^{ij} = \epsilon^{ijkl} G_{kl}$.

The equations of motion are the Einstein and Maxwell Equations which
yield\fnote{\star}{There are corrections to these expressions for $F_{ij}$
and $G_{ij}$, as is evident by taking exterior derivatives of both sides of
Eq.~(\ref{extder}). For a discussion, see~\cite{papa}. They do not affect our
results.}
\be
\psi = {\cal A}_0\quad,\quad F_{ij} = 2\psi\, {\cal F}_{ij}\quad,\quad
G_{ij} = 3\psi^2\, {\cal F}_{ij}
\ee
where ${\cal A}_\mu$ is the vector potential generated by the source current
$j^\mu$ in flat spacetime and ${\cal F}_{\mu\nu}$ is its field strength:
\be
\label{eq73}
\partial_\mu {\cal F}^{\mu\nu} = 2\pi^2\, j^\nu \quad,\quad
{\cal F}_{\mu\nu} = \partial_\mu {\cal A}_\nu - \partial_\nu {\cal A}_\mu
\quad,\quad j^\mu = \sum_I M_I\int dX_I^\mu \; \delta^5 (x-X_I)
\ee
Notice that $j^\mu$ is a gauge (reparametrization) invariant quantity, as it 
should be.
For the sources, we obtain the Lorentz force equation,
\be
\ddot X_I^\mu + \Gamma_{\nu\lambda}^\mu \, \dot X_I^\nu \dot X_I^\lambda
= \dot X_I^\nu F_{\;\;\nu}^\mu
\ee
The path integral is
\be
Z = {\cal N} \; \int {\cal D}g \, {\cal D} A \, \prod_I {\cal D} X_I\, {\cal 
D}P_I
\delta (\chi_I)\, e^{iS}
\ee
To calculate this, we need to fix the gauge in the sources. The simplest choice 
is
\be\label{eq72}
X_I^0 = t
\ee
for all black holes.
Then the current becomes
\be
j^\mu = \sum_I M_I v_I^\mu \delta^4 (\vec x - \vec X_I)\quad,\quad v_I^\mu
= (1, \vec v_I)\quad,\quad \vec v_I = {d\vec X_I\over dt}
\ee
where $|\vec v_I|\ll 1$ (non-relativisic limit).
The vector potential is found to be
\be
{\cal A}_0 = \psi = \sum_I {M_I\over (\vec x - \vec X_I(t))^2}\quad,\quad
\vec{\cal A} = \sum_I {M_I\vec v_I\over (\vec x - \vec X_I(t))^2}
\ee
After solving the constraints $\chi_I = 0$, in the non-relativistic limit,
\be
P_{I0} = {g^{ij} \pi_i\pi_j\over 2M_I} + M_IA_0
\ee
the action for the sources becomes
\be
S_{source} = \sum_I \int dt (\dot X_I^i P_{Ii} - H_I)\quad, \quad H_I = - P_{I0}
\ee
After integrating over the momenta $\vec P_I$ in the path integral, we obtain
\be
Z = {\cal N} \int \prod_I {\cal D} X_I\; e^{iS}
\ee
where $S = S_{fields} + S_{source}$, $S_{fields}$ is given by~(\ref{eq69})
and
\be
S_{source} = \sum_I M_I \int dt (\half \psi^2 \dot{\vec X_I^2} + \dot{\vec
X_I} \cdot \vec P)
\ee
After some algebra, the action can be cast into the form~\cite{bib4}
\be
S = S_{fields} + S_{source} = \half \int dt \sum_{I\ne J} G_{IJ}  (\vec v_I -
\vec v_J)^2 +\dots
\ee
where
\be G_{IJ} = G_{IJ} (\vec X_I - \vec X_J) = {M_IM_J(M_I+M_J)\over (\vec X_I-\vec 
X_J)^4}
\quad,\quad \vec v_I = {dX_I\over dt}\ee
and we have represented by dots the remaining less singular terms (they
involve three-point interactions).
This leads to a Hamiltonian with no well-defined ground state. This pathology is
shared with the case of a black hole moving in the background of another static
black hole. This is expected, because the latter case is encompassed by the
multi-black hole system. We need to be more careful in implementing the 
quantization
procedure as problems arise from the signularities of moduli space (when two 
black holes
become coincident).

Instead of collectively identifying all $X_I^0$ coordinates with time, we
shall adopt a gauge similar to the one discussed in the previous Section
({\em cf.}~Eq.~(\ref{gftan})),
\be
h_J (X_J^\mu) = t
\ee
To determine the function $h_J$ for the $J$th black hole, we work as
follows. Consider the $J$th black hole. It moves under the influence of
the (gravitational and electromagnetic) fields created by the other
black holes (given by Eqs.~(\ref{eq70})~-~(\ref{eq73}), where the sum in
(\ref{eq73}) runs over $I\ne J$). Of course, this statement is only valid in
the non-relativistic limit we are considering where the Einstein-Maxwell
Equations are essentially linearized. As our chosen black hole approaches
on of the other black holes (the $I$th one, say), the influence of the
rest of the system becomes negligible. By switching to the rest frame of
the $I$th black hole, the dynamics of our chosen ($J$th) black hole is
governed by the action discussed in the previous Section. Therefore, the
gauge-fixing condition $h_J = t$ should reduce to the gauge~(\ref{gftan})
in that frame and in the limit where the $I$th and $J$th black holes
become coincident. As discussed in the previous Section, we expect that
the net effect in the non-relativistic limit will be the addition of
a potential of the form~({\em cf.}~Eq.~(\ref{eq60}))
\be
K_I^{(J)} = {M_JM_I^2\over 2(\vec X_J - \vec X_I)^2}
\ee
Switching back to the center-of-mass frame from the rest frame of the
$I$th black hole does not alter this conclusion, because we need only
perform a Galilean transformation in the non-relativistic limit.
Repeating the argument with the rest of the black holes in the system,
we expect to obtain a net additional potential
\be
K^{(J)} = \sum_{I\ne J} K_I^{(J)} = {M_J\over 2} \sum_{I\ne J}
{M_I^2\over (\vec X_J - \vec X_I)^2}
\ee
To implement the above considerations in detail, let us introduce the
coordinates~({\em cf.}~Eq.~(\ref{eq34}))
\be\label{eqB}
x_I^{(J)\pm} = X_J^0 \pm {M_I^{3/2}\over 2(\vec X_J - \vec X_I)^2}
\ee
and the gauge-fixing condition~({\em cf.}~Eq.~(\ref{gftan}) where we set
$\omega =1/2$, for simplicity)
\be\label{newg}
h_J (X_J^\mu) = X_J^0 + \sum_{I\ne J} \left(
\arctan \left\{ {X_J^0\over 1+\halfhalf
x_I^{(J)+} x_I^{(J)-}} \right\} -X_J^0 \right) =t
\ee
Notice that as the distance $(\vec X_J - \vec X_I)^2 \to 0$ for a fixed
$I$, with all other distances remaining finite, the above definitions
coincide with Eqs.~(\ref{eq34}) and (\ref{gftan}), respectively. This gauge choice is
guaranteed to give no boundary contribution, because $\dot h_J \to 0$
near the boundary of moduli space. As in the previous Section, in order to calculate the non-relativistic limit, we need to augment
the Lagrangian by adding a total time derivative (which again gives no
contribution at the boundary) ensuring that the Lagrangian will have no
explicit dependence on $h_J$ (leading to a time independent Hamiltonian).
Thus, we define the action for the $J$th black hole by~({\em cf.}~Eq.~(\ref{eq34}))
\be\label{eqL}
S_J = \int dt \dot X_J^\mu P_{J\mu} + \dot \Lambda^{(J)}\quad,\quad
\Lambda^{(J)} = \sum_{I\ne J} \Lambda_I^{(J)}\quad,\quad
\Lambda_I^{(J)} = - {M_J\sqrt{M_I}\over 4}\; \ln \left\{ {1+\halfhalf (x_I^{(J)+})^2
\over 1+\halfhalf (x_I^{(J)-})^2}\right\}
\ee
It is easily verified that in the non-relativistic limit,
\be\label{eqC1}
t= h_J \approx X_J^0 \quad,\quad \Lambda_I^{(J)} \approx X_J^0 {M_JM_I^2\over
8(\vec X_J - \vec X_I)^2} = \halfhalf tK_I^{(J)}
\ee
Therefore in the non-relativistic limit, the additional term in the action reads
\be\label{eqC2}
\dot \Lambda^{(J)} \approx \halfhalf \sum_{I\ne J} K_I^{(J)} = \halfhalf K^{(J)}
\ee
in agreement with our expectations (having set $\omega = \half$).

The above gauge-fixing procedure can be repeated with the rest of the black holes
in the system. We modify the action for the sources
by a total time derivative which does not alter the dynamics, but simplifies the
calculation of the non-relativistic limit. Thus, we define the action for the
sources by~({\em cf.}~Eq.~(\ref{eq65}))
\be
S_{source} = \sum_J S_J = \sum_J \int dt \left(\dot X_J^\mu P_{J\mu} + \dot \Lambda^{(J)} \right)
\ee
where $\Lambda^{(J)}$ is given by~(\ref{eqL}). In the non-relativistic limit,
the net effect of the gauge~(\ref{newg}) is the addition of the potential $\halfhalf K$, where
\be\label{eqK}
K = \sum_J K^{(J)} = \sum_J \sum_{I\ne J} {M_JM_I^2\over
2(\vec X_J - \vec X_I)^2} = \sum_{I< J} {M_IM_J(M_J+M_I)\over
2(\vec X_J - \vec X_I)^2}
\ee
In conclusion, we have shown how to fix the gauge in a way that no obstruction
occurs from the singularities of moduli space (when two black holes approach
each other). The resultant Hamiltonian differs from the one obtained in the
na\"\i ve gauge~(\ref{eq72}) by the addition of the potential~(\ref{eqK}).
This is accordance with the DFF prescription~\cite{bib1}.

The na\"\i ve gauge~(\ref{eq72}) suffers from an obstruction at the
singularities of moduli space. Once the obstruction is correctly accounted
for, the resulting theory is equivalent to the one in which the Hamiltonian is
modified by the addition of the potential $K$~(Eq.~(\ref{eqK})). This is
because the underlying theory is gauge-invariant.

\section{Four spacetime dimensions}
\label{sec4D}

%\section{Four-dimensional Reissner-Nordstr\"om black hole}

The four-dimensional case was originally discussed by Ferrel and
Eardley~\cite{bib5}. The results are similar to the five-dimensional case.
Therefore, we will only summarize the major differences between the two cases.

In four dimensions, the metric due to an extreme Reissner-Nordstr\"om black hole 
is
\be
ds^2 = - {1\over \psi^2} dt^2 + \psi^2 d\vec x^2 \quad,\quad \psi = 1 +
{M\over |\vec x|}
\ee
and the vector potential is
\be
A_t = {1\over\psi} -1 \quad,\quad \vec A = 0
\ee
where the vectors live in a three-dimensional Euclidean space.
Near the horizon, $\psi = M/|\vec x|$. Using polar coordinates and switching
variables to $\psi$, we obtain
\be
ds^2 = -{1\over\psi^2} \left( dt^2 - M^2 d\psi^2\right)
+ M^2 d\Omega_2^2
\ee
Defining
\be
x^\pm = t \pm M \; \psi
\ee
the metric becomes
\be
ds^2 = - {1\over \psi^2} \; dx^+ dx^- + M^2 d\Omega_2^2
\ee
This is of the same form as in five spacetime dimensions (Eq.~(\ref{metr4})),
apart from the scale factor in the spherical part of the metric. Working as in
Section~\ref{sec4}, we arrive at an $SL(2,{\bf R})$ conformal algebra consisting
of the operators
\be
H = {|\vec x|^3 \vec P^2\over 2mM^3} \quad, \quad D = -2\vec x\cdot \vec P
\quad,\quad K = {2mM^3\over |\vec x|}
\ee
in the non-relativistic limit. In the gauge
\be\label{4dg}
h(x^+, x^-) = \arctan\left( {\omega x^++\omega x^-\over 1-\omega^2
x^+x^-} \right) = \tau
\ee
the Hamiltonian becomes
\be
{\cal H} = {1\over 2\omega} (H+\omega^2 K)
\ee
which has a well-defined vacuum state. The system reduces to a harmonic 
oscillator
if we change variables to $u = M\sqrt\psi$.

As Ferrel and Eardley
showed~\cite{bib5}, in the case of two slowly-moving black holes,
in the center-of-mass frame and in the near-horizon limit,
the Hamiltonian becomes
\be
H = {|\vec X_2-\vec X_1|^3 (\vec P_2 - \vec P_1)^2\over 2\mu (1-2\mu /M)M^3} = 
{|\vec X_2- \vec X_1|^3 (\vec P_2 - \vec P_1)^2\over 2
M_1M_2(M_1^2+M_2^2)}
\ee
where $\mu = M_1M_2/(M_1+M_2)$ is the reduced mass and $M=(M_1+M_2)$ is the 
total
mass of the system. This generalizes to an arbitrary number of maximally-charged
slowly-moving black holes,
\be
H = \sum_{I\ne J}
{|\vec X_I- \vec X_J|^3 (\vec P_2 - \vec P_1)^2\over 2
M_IM_J(M_I^2+M_J^2)} + \dots
\ee
where we have omitted the less singular terms. Evidently, this system has
no ground state. The origin of this pathology is the same as in the five-dimensional case.
To remedy this, we work as in Section~\ref{sec4a}.
The action for the fields is now~({\em cf.}~Eq.~(\ref{eqA}))
\be
S_{fields} = {1\over 16\pi} \int d^4x \, \sqrt g (R-F^2)
\ee
whereas the action for the sources remains unchanged.
We introduce the gauge-fixing condition~(\ref{newg}), where~({\em cf.}~Eq.~(\ref{eqB}))
\be
x_I^{(J)\pm} = X_J^0 \pm {M_I^2\over 2|\vec X_J - \vec X_I|}
\ee
and augment the Lagrangian for the sources with the total derivative~({\em cf.}~Eq.~(\ref{eqL}))
\be
\dot\Lambda = \sum_J \dot\Lambda^{(J)}\quad,\quad \Lambda^{(J)} = \sum_{I\ne J} \Lambda_I^{(J)}
\quad,\quad
\Lambda_I^{(J)} = - 2M_J M_I \ln\left\{ {1+\halfhalf (x_I^{(J)+})^2
\over 1+\halfhalf (x_I^{(J)-})^2} \right\}
\ee
In the non-relativistic limit, the gauge condition~(\ref{newg}) reduces to
$h_J\approx X_J^0 = t$ and the additional term in the action to~({\em cf.}~Eqs.~(\ref{eqC1}) and (\ref{eqC2}))
\be
\dot\Lambda \approx \halfhalf \sum_{I\ne J} K_I^{(J)}
\ee
where
\be
K_I^{(J)} = {2M_JM_I^3\over |\vec X_J - \vec X_I|}
\ee
Therefore, in the gauge~(\ref{newg}), the Hamiltonian is
modified by the potential $K = \sum_{I< J} K_{IJ}$, where~({\em cf.}~Eq.~(\ref{eqK}))
\be
K_{IJ} = {M_IM_J(M_I^2+M_J^2)\over 2|\vec X_I - \vec X_J|}
\ee
similar to five
spacetime dimensions~(Section~\ref{sec4a}) and in accordance with the DFF
prescription~\cite{bib1}.

\section{Conclusions}
\label{sec5}

We considered the problem of quantization of a system of slowly-moving
extreme Reissner-Nordstr\"om black holes. The moduli have dynamics governed by
(super)conformal quantum mechanics and the Hamiltonian has no well-defined
ground state. This problem is fixed by an application of the DFF
trick~\cite{bib1}. To justify this trick on physical grounds, we approached the
problem through the path integral and the Faddeev-Popov gauge-fixing
procedure. We showed that the DFF trick can be understood in terms of the
standard Faddeev-Popov procedure~\cite{bib7}.  We started with a discussion
of the quantization of a particle in the presence of a background metric
field as well as an external vector potential. We performed the standard
Faddeev-Popov procedure in the canonical formalism and showed its connection
to commutation relations through Dirac brackets. We then applied the procedure
to the case of an extreme Reissner-Nordstr\"om black hole~\cite{bib4}. We
showed that the na\"\i ve identification of time coordinate (which leads to a
Hamiltonian system with no well-defined ground state) corresponds to a gauge
which is not ``good." We found that in this gauge the Faddeev-Popov procedure
encounters an obstruction at the boundary of spacetime introducing an
additional constraint there. This alters the standard commutation relations
and the eigenvalue problem for the attendant Hamiltonian system. We did not
calculate the effects of this obstruction explicitly. This would require the
introduction of a regulator which would break gauge invariance explicitly and
therefore alter the commutation rules. Instead, we exhibited another set of
gauges where no obstruction existed on the boundary. We showed that this set
of gauges led to a Hamiltonian system with a well-defined vacuum, equivalent
to the one obtained through the DFF trick~\cite{bib1}.

We then applied our procedure to multiple black hole scattering~\cite{bib4}.
We noted that the underlying theory
is a gauge theory, so the Faddeev-Popov procedure should be applicable. We
discussed a systematic implementation of the quantization procedure which
correctly accounted for the singularities of moduli space. Each black hole is
described by moduli (position vector) $X^\mu (\tau)$ and the action is
reparametrization invariant. This gauge invariance necessitated the
introduction of gauge-fixing conditions equal in number to the number of black
holes. By identifying $X^0$ with time for all black holes, one arrives at the
standard  Hamiltonian that possesses no well-defined ground state. This
pathology comes from a subtlety in the Faddeev-Popov quantization procedure
that does not take into account the singularities of moduli space. To properly
account for these singularities would be tedious (entailing the introduction
of a regulator) and would lead to a modification of the quantization rules.
Instead, we introduced gauge-fixing conditions that did not suffer from this
pathology. The resultant Hamiltonian differed from the pathological one by
the addition of the potential $\halfhalf K$, where $K$ is the
generator of special conformal
transformations, in accordance with the DFF prescription.

Our method is generalizable to any system of black holes and more general
solutions of the Einstein-Maxwell equations. It would be interesting to apply
the Faddeev-Popov procedure to these systems, such as near-extreme black
holes. This would enable us to move away from the AdS limit.

\end{document}